\documentclass[reprint,aps,prb,amsmath,amssymb,longbibliography, notitlepage]{revtex4-1}
\usepackage{bm, graphicx}
\usepackage[stable]{footmisc}
\usepackage{physics}
\usepackage{multirow}
\usepackage{tabularx}
\usepackage[colorlinks, linkcolor= blue, citecolor = blue, urlcolor=blue]{hyperref}
\usepackage{array}
\usepackage{pifont}

\newcolumntype{P}[1]{>{\centering\arraybackslash}p{#1}}
{
\newcommand{\cmark}{\ding{51}} % Check mark
\newcommand{\xmark}{\ding{55}} %cross mark

\def\nn{\nonumber}
\def\bea{\begin{eqnarray}}
\def\eea{\end{eqnarray}}
\def\be{\begin{equation}}
\def\ee{\end{equation}}

\def\tc{\textcolor}

\def\kb{{\bm k}}

\def\e{\varepsilon}
\def\m{\mathcal}
\def\bal{\begin{aligned}}
\def\eal{\end{aligned}}
\def\bseq{\begin{subequations}}
\def\eseq{\end{subequations}}
\usepackage{braket}

\begin{document}

%\title{Planar Nernst Effect In Quasi 2D-materials}
\title{Planar Nernst effect from hidden band geometry in layered two-dimensional materials}
\author{Rahul Biswas}
\thanks{R. B. and H. V. contributed equally to the manuscript}
\affiliation{Department of Physics, Indian Institute of Technology, Kanpur-208016, India.}
\author{Harsh Varshney}
\thanks{R. B. and H. V. contributed equally to the manuscript}
\affiliation{Department of Physics, Indian Institute of Technology, Kanpur-208016, India.}
%\emial{}
\author{Amit Agarwal}
\email{amitag@iitk.ac.in}
\affiliation{Department of Physics, Indian Institute of Technology, Kanpur-208016, India.}
\begin{abstract}
The Nernst effect is a versatile phenomenon relevant for energy harvesting, magnetic sensing, probing band topology, and charge-neutral excitations. The planar Nernst effect (PNE) generates an in-plane voltage transverse to an applied temperature gradient under an in-plane magnetic field. Conventional Berry curvature-induced PNE is absent in two-dimensional (2D) systems, as the out-of-plane Berry curvature does not couple to the in-plane electron velocity. Here, we challenge this notion by demonstrating a distinct planar Nernst effect in quasi-2D materials (2DPNE). We show that the 2DPNE originates from previously overlooked planar components of Berry curvature and orbital magnetic moment, arising from inter-layer tunneling in multilayered 2D systems. We comprehensively analyze the band-geometric origin and crystalline symmetry constraints on 2DPNE responses. We illustrate its experimental feasibility in strained bilayer graphene. Our findings significantly expand the theoretical understanding of planar Nernst effects, providing a clear pathway for next-generation magnetic sensing and energy-harvesting applications.
\end{abstract}

\maketitle

%%%%%%%%%%%%%%%%%%%%%%%  Introduction  %%%%%%%%%%%%%%%%%%
\section{Introduction}

The planar Nernst effect (PNE) is a thermoelectric phenomenon generating longitudinal and transverse electrical currents within the plane defined by an applied temperature gradient and magnetic field. It emerges from coupling Berry curvature (BC) and orbital magnetic moment (OMM) with in-plane electron velocity components~\cite{avery2012observation, sharma2016planar, das2019berry, varshney2025intrinsic, bui2014anomalous, wesenberg2018relation,sharma2019transverse, pena2022seebeck, dubey2024magnetically}.
This phenomenon enables pure spin current generation without an accompanying charge current, vital for spintronic applications~\cite{boona2014spin,bose2019recent,liu2023absence, reimer2017quantitive}, and efficiently converts temperature gradients into electrical currents, essential for energy harvesting technologies~\cite{Snyder2009thermoelectric,sothmann2014thermoelectric,yu2024ambient, chhatrasal2016, massetti2021unconventional, khanh2025gapped}. Its sensitivity to magnetic field orientation makes it ideal for precision magnetic and heat flux sensors~\cite{pandey2024anomalous,martens2018anomalous,chen2022large,tanaka2023roll}. Moreover, PNE serves as a powerful probe of band geometry in topological semimetals and insulators~\cite{xu2019large,ceccardi2023anomalous, pan2024topological}.

Despite these promising applications, conventional PNE is fundamentally limited in two-dimensional (2D) systems. In an ideal 2D system, electron motion is confined strictly to the plane, forcing the Berry curvature (BC) and orbital magnetic moment (OMM) to have only out-of-plane components. These out-of-plane components do not couple to the in-plane electron velocities or magnetic fields, resulting in the complete suppression of conventional PNE in 2D systems. Consequently, a broad class of novel 2D materials remains unexplored for potential PNE applications. Recently, 
we addressed a similar limitation in the planar Hall effect (PHE) for 2D materials, demonstrating a unique PHE enabled by the hidden in-plane BC and OMM components in layered 2D systems arising from interlayer coupling in layered 2D systems~\cite{ghorai2025planar}. This prompts a natural question: can similar mechanisms support two-dimensional PNE  in layered heterostructures? If feasible, this would unlock numerous materials, such as multilayered graphene, transition metal dichalcogenides (TMDs), and other van der Waals heterostructures, for innovative thermoelectric, spintronic, and energy-harvesting technologies. 

%%%%%%%%%%%%%%%%%%%%%%%%%%%%%%%%%%%%%%%
\begin{figure}[t]
    \centering
    \includegraphics[width=\linewidth]{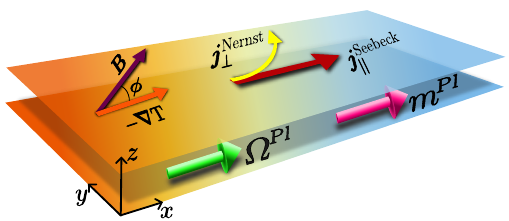}
    \caption{{\bf Schematic for measuring 2D Planar Nernst Effect (2DPNE) in layered 2D materials.} Finite interlayer tunneling in layered materials enables the planar Berry curvature $(\Omega^{\rm Pl})$ and planar orbital magnetic moment ($m^{\rm Pl}$). These couple to an in-plane magnetic ($B$) field and temperature gradient to generate longitudinal planar Seebeck (${\bm j}^{\rm Seebeck}_{||}$) and transverse planar Nernst (${\bm j}^{\rm Nernst}_{\perp}$) thermoelectric currents.}
    \label{Fig1}
\end{figure}
%%%%%%%%%%%%%%%%%%%%%%%%%%%%%%%%%%%%%%

In this paper, we introduce a unique two-dimensional planar Nernst effect (2DPNE) in layered materials, enabled by previously overlooked planar band geometric quantities. Unlike ideal 2D systems, layered materials, with finite interlayer coupling along with broken space-inversion or time-reversal symmetry, can host in-plane components of BC and OMM~\cite{drigo2020chern,hara2020current,kim2021vertical}. These hidden planar BC (${\bm \Omega}^{\rm Pl}$) and planar OMM (${\bm m}^{\rm Pl}$) contributions play a crucial role in enabling 2DPNE. We provide a detailed analysis of their origin and the resulting 2DPNE responses. Figure~\ref{Fig1} illustrates the  2DPNE setup, where an in-plane temperature gradient ($-{\bm \nabla} T$) and magnetic field (${\bm B}$) give rise to both the planar Seebeck (${\bm j}^{\rm Seebeck}_{||}$) and planar Nernst  (${\bm j}^{\rm Nernst}_{\perp}$) currents. We also identify the crystalline point-group symmetry requirements for finite 2DPNE and analyze its angular dependence on the angle $\phi$ between $-{\bm \nabla} T$ and ${\bm B}$. Using Bernal-stacked bilayer graphene (BLG) as a case study, we demonstrate a significant gate-tunable 2DPNE response. Our findings lay the foundation for further experimental exploration and future design of tunable thermoelectric and energy harvesting devices based on 2DPNE.

%%%%%%%%%%%%%%%%%%%% Table.1 : Highlighting our 2DPNE responses %%%%%%%%%%%%%%%%%%%%%%%%%%%%%%
\begin{table*}[t!]
\caption{We present the general expressions of the third ($\chi_{ab;c}$) and fourth-rank ($\chi_{ab;cd}$) 2DPNE response tensor. Here, we segregate these response tensors into Berry curvature (BC), orbital magnetic moment (OMM), and mixed (BC + OMM) contributions as discussed in Eq.~\eqref{eq:splitting}. For conciseness, we express these contributions as $\chi^{\rm BC/OMM/BC+OMM}_{ab;c(d)}  = \frac{e^2}{\hbar} \sum_n \int [d\kb] (A_1 f'_0 + A_2 f''_0 + A_3 f'''_0)  ~ + (c \leftrightarrow d) $, where $f'_0,~f''_0, ~{\rm and}~f'''_0$ denote the first, second, and third order derivatives of the Fermi Dirac distribution function ($f_0$) with respect to the energy, respectively. The prefactors $A_1, ~A_2, ~{\rm and}~ A_3$ encapsulate the dependence on the band velocity and the relevant planar band geometric quantities for each type of contribution. The symmetrizing term $(c \leftrightarrow d)$ symmetrizes $\chi_{ab;cd}$ in the magnetic field indices. For notational simplicity, we define ${\bar \e} = (\varepsilon-\mu)$, $\Omega_{V} = \bm v \cdot \bm \Omega^{\rm Pl}$, and use $\int[d\kb] = \int \frac{d^d\kb}{(2\pi)^d}$ to denote the $d$-dimensional momentum-space integration.}
\centering
\begin{tabular}{P{1cm} P{1cm} P{5cm} P{9.5cm}}
\hline\hline
    
      \multicolumn{4}{c}{Third and Fourth rank 2DPNE response Tensors } \\ 
     \multicolumn{2}{c}{Band geometric origin} &$\chi_{ab;c}$ & $\chi_{ab;cd}$ \\
     \hline 
     % Berry curvature components
     \multirow{3}{6em}{BC} & $A_1$ & $ e {\bar \e}(v_{a}v_{b}\Omega^{\rm Pl}_{c} -(v_{a}\delta_{bc}+v_{b} \delta_{ac}) \Omega_{V}) $ & $
 \frac{e^{2}}{2\hbar} {\bar \e} \bigg((v_{a}\delta_{bd}+v_{b}\delta_{ad})\Omega^{\rm Pl}_{c}\Omega_{V} -v_{a}v_{b}\Omega^{\rm Pl}_{c}\Omega^{\rm Pl}_{d} - \Omega^2_{V}\bigg)  $ \\ 
     & $A_2$ & 0 & 0 \\
     & $A_3$ & 0 & 0 \\
     \hline 
     % OMM components
     \multirow{3}{6em}{OMM} & $A_1$ & ${\bar \e}(v_{a}\partial_{b}m^{\rm Pl}_{c}+v_{b}\partial_{a}m^{\rm Pl}_{c}) + \hbar v_{a}v_{b}m^{\rm Pl}_{c} $ & $-\frac{1}{2 \hbar}\bigg( {\bar \e} \partial_{a}m^{\rm Pl}_{c} \partial_{b}m^{\rm Pl}_{d}  + \hbar(v_{a}\partial_{b} m^{\rm Pl}_{c} +  v_{b}\partial_{a} m^{\rm Pl}_{c})m^{\rm Pl}_{d} \bigg)$ \\
     & $A_2$ & $\hbar{\bar \e} v_{a}v_{b}m^{\rm Pl}_{c} $ & $-\frac{1}{2}\bigg( {\bar \e}(v_{a}\partial_{b} m^{\rm Pl}_{c} +v_{b}\partial_{a}m^{\rm Pl}_{c} ) m^{\rm Pl}_{d}  + \hbar v_{a}v_{b}m^{\rm Pl}_{c} m^{\rm Pl}_{d}\bigg)$ \\
     & $A_3$ & 0 & $-\frac{\hbar {\bar \e}}{2}v_{a}v_{b} m^{\rm Pl}_{c} m^{\rm Pl}_{d}$   \\
     \hline 
     % Berry curvature + OMM components
       \multirow{3}{6em}{ BC + \\ OMM} & $A_1$  & 0 & $ \frac{e}{2 \hbar} \bigg[ {\bar \e} \bigg( (\partial_{a} m^{\rm Pl}_{c} \delta_{bd}+\partial_{b} m^{\rm Pl}_{c} \delta_{ad}) \Omega_{V} + (v_{a}\delta_{bc}+v_{b}\delta_{ac}) (\bm{\Omega}^{\rm Pl}\cdot \bm{\nabla}_{\bm k}m^{\rm Pl}_{d}) -(v_{a}\partial_{b} m^{\rm Pl}_{d} + v_{b}\partial_{a} m^{\rm Pl}_{d})\Omega^{\rm Pl}_{c} \bigg) +\hbar \bigg((v_{a}\delta_{bc}+v_{b}\delta_{ac})\Omega_{V}  -  v_{a}v_{b}\Omega^{\rm Pl}_{c} \bigg) m^{\rm Pl}_{d}\bigg]   $ \\ 
     & $A_2$  & 0 & $ \frac{e}{2} {\bar \e} \left((v_{a}\delta_{bc}+v_{b}\delta_{ac})\Omega_{V}- v_{a}v_{b}\Omega^{\rm Pl}_{c} \right) m^{\rm Pl}_{d}$ \\
   & $A_3$  & 0 & 0 \\
     \hline 
     \hline
\end{tabular}
\label{tab1}    
\end{table*}
%%%%%%%%%%%%%%%%%%%%%%%%%%%%%%%%%%%%%%%%%%%%%%%%%%%%%%%%%%%%%%%%%%%%%%%%%%%%%%%%%%%%%%%%%%%%%%%
%

%%%%%%%%%%%%%%%%%%%%%%%  Section 2  %%%%%%%%%%%%%%%%%%
\section{Planar Berry curvature and orbital magnetic moment }

In this section, we define and motivate the concept of planar BC and OMM, which play a central role in enabling the 2DPNE in layered systems. In ideal 2D materials, electron motion is strictly confined to the $x$–$y$ plane, and as a result, the BC and OMM vectors possess only out-of-plane components (along $\hat{z}$). However, this constraint is relaxed in quasi-2D systems composed of two or more coupled atomic layers. Interlayer tunneling in such systems induces out-of-plane electronic delocalization, which can generate finite in-plane components of the BC and OMM. We refer to these as the planar BC and planar OMM, and they are essential for realizing 2DPNE.

Conventionally, the BC and OMM of the $n$-th band are defined as~\cite{xiao2010berry,Chang_Niu_OMM, atencia2024orbital}
\begin{align*}
{\bm \Omega}_n (\kb) &= {\bm \nabla}_{\kb} \times {\bm {\m A}}_n(\kb), \\
{\bm m}_n (\kb) &= \frac{ie}{2\hbar} \bra{{\bm \nabla}_\kb u_{\kb}^n} \times (\e^n_\kb - {\m H}_0)\ket{{\bm \nabla}_\kb u_{\kb}^n},
\end{align*}
where ${\bm {\m A}}_n = \bra{u_\kb^n} i{\bm \nabla}_\kb \ket{u^n_\kb}$ is the Berry connection, $\e^n_\kb$ is the band energy, and $\ket{u_\kb^n}$ is the periodic part of the Bloch wavefunction for a system governed by Hamiltonian ${\m H}_0$.

In quasi-2D materials, these definitions have to be extended to incorporate the effects of interlayer tunneling. The resulting planar components of the BC and OMM are given by, 
\begin{subequations}\label{eq:planar_BC_OMM} 
\begin{align}
{\bm \Omega}^{\rm Pl}_{n \bm k} &= 2\hbar \sum_{n' \neq n}\frac{\text{Re}(\bm v_{nn^{'}} \times {\bm {\m Z}}_{n'n})}{\varepsilon^n_{\kb}- \varepsilon^{n'}_{\kb}}~,\label{Omega_pl} \\
{\bm m}^{\rm Pl}_{n \bm k} &= e\sum_{n' \neq n} \text{Re}(\bm v_{n n'} \times {\bm {\m Z}}_{n'n})~,\label{m_pl}
\end{align}
\end{subequations}
where the superscript `Pl' denotes the planar (in-plane) contributions. Here, 
$\hbar {\bm v}_{nn'}=\bra{u^n_{\bm k}} {\bm \nabla}_{\bm k}\mathcal{H}\ket{u^{n'}_{\bm k}}$ is the band velocity matrix element, and ${\bm {\m Z}}_{n'n} = \hat{\bm z} \bra{u^{n'}_\kb}\hat{\m Z}\ket{u^n_\kb}$ is the matrix element of the out-of-plane position operator $\hat{\m Z}$. These expressions reflect the geometric origin of planar BC and OMM arising from the interlayer tunneling. The emergence of ${\bm \Omega}^{\rm Pl}_{n\bm k}$ and ${\bm m}^{\rm Pl}_{n\bm k}$ is unique to layered materials with finite off-diagonal matrix elements of ${\bm {\m Z}}_{n'n}$ enabled by wavefunction overlaps between adjacent layers. These planar band geometric quantities vanish in monolayers with zero thickness or in multilayered systems with completely decoupled layers. See Appendix~\ref{planar_BC} for a detailed derivation. 

%%%%%%%%%%%%%%  section 3   %%%%%%%%%%%%%%%%%%%
\section{Theory of planar Nernst effect}

In this section, we develop a general theory of the planar Nernst effect within the semiclassical framework of charge transport. In this approach, electrons are described as wave-packets undergoing translational and rotational motion in momentum space under external perturbations. These dynamics generate two distinct contributions to the local current, a transported current arising from translational motion and a circulating (magnetization) current associated with the self-rotation of the wave-packet. Since the magnetization current circulates locally and does not contribute to the measurable current in transport experiments, the transported current constitutes the observable response.

The total local and magnetization current densities for a single band in the presence of a temperature gradient and magnetic field are given by~\cite{xiao2006berry, xiao2010berry}:
\bseq
\bea
{\bm j}^{\rm loc} &=&\int[d\kb] D_\kb^{-1} \left( -e\bm{\dot{r}} + {\bm \nabla}_{\bm r} \times {\bm m}_\kb \right) f({\bm r}, \kb)~, \\
{\bm j}^{\rm mag} &=& {\bm \nabla}_{\bm r} \times {\bm M}(\bm r)~.
\eea
\eseq
Here, $-e$ ($e>0$) is the electron's charge, $D_\kb = \left[1 + \frac{e}{\hbar} ({\bm B} \cdot {\bm \Omega}_\kb)\right]^{-1}$ is the modified phase space factor in presence of a magnetic field ${\bm B}$, ${\bm \Omega}_\kb$ is the Berry curvature, $\bm{\dot{r}}$ is the wave-packet velocity, and $f({\bm r}, \kb)$ is the non-equilibrium distribution function. The OMM of the wave-packet is denoted by ${\bm m}_\kb$, and ${\bm M}(\bm r)$ is the equilibrium orbital magnetization density. We define $\int [d\kb] = \int d^d\kb/(2\pi)^d$. The equilibrium magnetization density is given by~\cite{xiao2006berry, das2021intrinsic} 
\bea
{\bm M}({\bm r}) &=& \int [d\kb] D^{-1} f({\bm r}, \kb) {\bm m}_\kb~ \\ 
&&+  \beta^{-1} \int[d\kb] e/\hbar \ {\bm \Omega}_\kb \log{(1+e^{-\beta(\tilde{\varepsilon}_\kb-\mu)})}~, \nn 
\eea
where $\beta^{-1} = k_B T $ with $k_B$ and $T$ being the Boltzmann constant and the system temperature, respectively. The parameter ${\tilde \varepsilon}_\kb = \varepsilon_\kb - {\bm m}_\kb \cdot {\bm B}$ is the OMM modified band energy.

To obtain the measurable transported current, we subtract the magnetization current density from the local current density, ${\bm j} = {\bm j}^{\rm loc} - {\bm j}^{\rm mag} $, to obtain the physically relevant transported current density, 
\bea\label{eq:j_trans} 
{\bm j} &=& -e\int[d\kb] D_\kb^{-1} \bm{\dot{r}} f({\bm r}, \kb)~ \\ 
&& - \frac{e}{\hbar} \gradient_{\bm r} \times  \int[d\kb] \frac{1}{\beta} {\bm \Omega}_\kb \log{(1+e^{-\beta(\tilde{\varepsilon}_\kb-\mu)})}~. \nn 
\eea 
This is the expression we use to compute the planar Nernst current. To proceed, we determine the wave-packet velocity and the distribution function in the presence of a temperature gradient and magnetic field using the semiclassical Boltzmann equation. For planar currents, we restrict the Berry curvature ${\bm \Omega}_\kb$ and OMM ${\bm m}_\kb$ to their in-plane components.

Treating the external fields perturbatively, we derive the nonequilibrium distribution function (see Eq.~\eqref{A4} in Appendix~\ref{2DPNE_current_density}) and obtain the planar Nernst current density. We retain temrs which are linear in the temperature gradient and up to second order in the magnetic field strength $B$. The resulting generic form of the planar Nernst current is given by, 
\be\label{eq:j_cur}
 j_{a} = \frac{\tau }{eT} \left[ \chi_{ab;c}~B_c \nabla_bT +  \chi_{ab;cd}~B_c B_d \nabla_b {T} \right]~.
\ee 
Here, $\tau$ is the electron scattering time, $\{a,~b,~c,~d \} \in \{ x,~y \}$ represent the 2D Cartesian coordinate, and $\nabla_b T$ is  the temperature gradient along $b-$direction. The Einstein summation convention over repeated spatial indices is implied. Here, $\chi_{ab;c}$ and $\chi_{ab;cd}$ are the third- and fourth-rank response tensors characterizing the 2DPNE. See Appendix.~\ref{2DPNE_current_density} for the detailed derivations and calculations. Further, we decompose the 2DPNE responses ($\chi_{ab;c}$ and $\chi_{ab;cd}$) as a sum of planar BC, planar OMM, and mixed terms as 
\be \label{eq:splitting}
\chi_{ab;c(d)} = \chi_{ab;c(d)}^{\rm BC} + \chi_{ab;c(d)}^{\rm OMM} + \chi_{ab;c(d)}^{\rm BC + OMM}~.
\ee 
Here, $\chi^{\rm BC}$ is the 2DPNE response tensor stemming solely from the planar Berry curvature, $\chi^{\rm OMM}$ accounts for the OMM-induced 2DPNE contribution, while $\chi^{\rm BC+OMM}$ captures the 2DPNE response arising from both BC and OMM. 
These individual components are tabulated in Table~\ref{tab1} and discussed in detail in Appendix~\ref{2DPNE_current_density}. 
These expressions completely describe all contributions to the longitudinal and transverse 2DPNE responses and form the main result of our paper.

%We present the detailed expression of these contributions in Table~\ref{tab1}. 

%%%%%%%%%%%%%%%%%%%% figure 2%%%%%%%%%%%
\begin{figure}[t!]
    \centering
    \includegraphics[width=\linewidth]{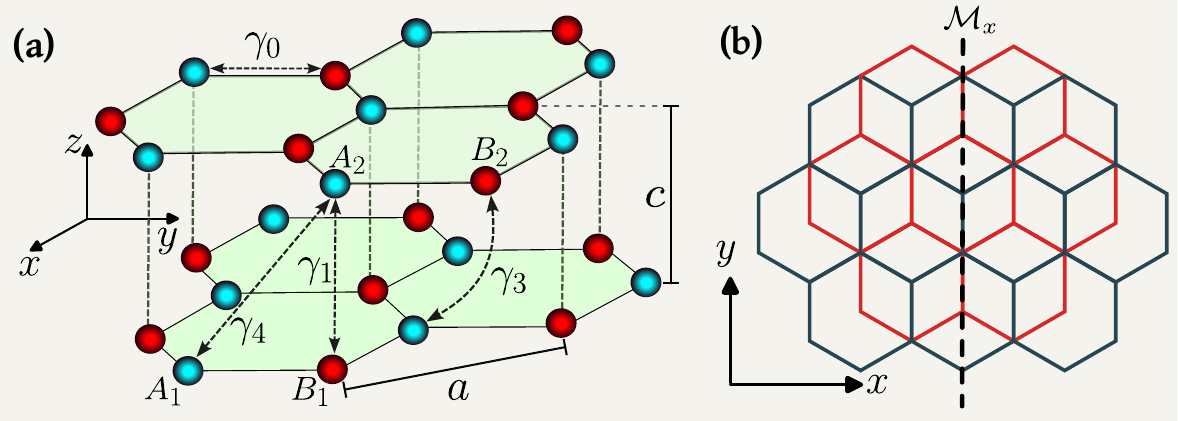}
    \caption{\textbf{Lattice structure and mirror symmetry of Bernal stacked bilayer graphene.} (a) Bernal stacked bilayer graphene's lattice structure with an interlayer distance $c$ and lattice constant $a$. Different intra- and inter-layer hoppings are depicted as  $\gamma_0,~\gamma_1,~\gamma_3,$ and $\gamma_4$. (b) The top view of the bilayer graphene displays its zigzag and armchair directions along the $x$ and $y$ axes. Bilayer graphene has a mirror symmetry about its armchair direction ($y$ axis in this figure). We represent this mirror symmetry by ${\cal M}_x$ and it transform the coordinate $(x,y)$ as ${\cal M}_x :(x,y) \to (-x, y)$. }
    \label{fig:lattice}
\end{figure}
%%%%%%%%%%%%%%%%%%%%%%%%%%%%%%%%
%%%%%%%%%%%5

%%%%%%%%%%%% fig 3 %%%%%%%%%%%%%%%%%%
\begin{figure*}[t!]
    \centering
    \includegraphics[width=.9\linewidth]{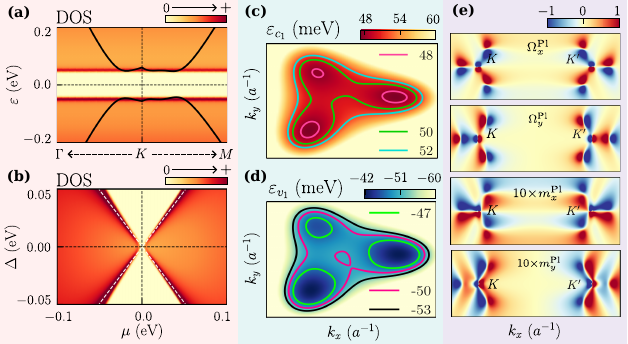}
    \caption{\textbf{Band structure, Density of states, Lifshitz transitions, and band geometric quantities of strained bilayer graphene. } (a) The Low-energy electronic band-dispersion (black lines) of the strained BLG near one of the Dirac points obtained from the tight-binding Hamiltonian in Eq.~\eqref{hamiltonian}. The background color displays the Density of states (DOS) as a function of energy $\varepsilon$. The DOS has pronounced peaks, or van Hove singularity (VHS), near the band edges due to the presence of saddle points in the band dispersion. (b) The variation of the DOS as a function of both chemical potential $\mu$ and the interlayer potential $\Delta$, which can be controlled via the gate voltages. The white dashed lines denote the evolution of the VHS point in the $\mu-\Delta$ space. In (c) and (d), we highlight the Lifshitz transition by showing a sudden change in the Fermi surface topology near the Dirac point. In our strained BLG system, these Lifshitz transitions occur around $50~$meV. (e) The upper (lower) two panels present the momentum space distribution of the $x$ and $y$ components of the planar BC (OMM) for the first conduction band in units of $a^2~(\frac{e}{\hbar}a^2 \cdot$V). We used the model parameters discussed in Sec.~\ref{sec:strBLG}. In all panels except (b), we have taken the interlayer potential $\Delta = 0.05~$eV.}
    \label{fig3}
\end{figure*}
%%%%%%%%%%%%%%%%%%%%%%%%%%%%

%%%%%%%%%%%%%%%%%%%%%%%%%%%
\section{Symmetry restriction on 2DPNE responses}
In this section, we analyze the restrictions imposed by both fundamental symmetries of space-inversion (${\cal P}$) and time-reversal (${\cal T}$), as well as crystalline point-group symmetries on the planar Nernst response tensors. The general form of the 2DPNE current, linear in the applied temperature gradient, is given in Eq.~\eqref{eq:j_cur}. Under ${\cal P}$ symmetry, both the current ${\bm j}$ and the temperature gradient ${\bm \nabla} T$ reverse sign, while $\tau$ and ${\bm B}$ remain invariant. Consequently, ${\cal P}$ symmetry imposes no constraint on the response tensors $\chi_{ab;c}$ and $\chi_{ab;cd}$.

%%%%%%%%%%%%%% table 2: Point group symmetry table %%%%%%%%
\begin{table*}[t!]
\setlength{\tabcolsep}{6.5pt}
\caption{ The symmetry restrictions of the B-linear and B-quadratic 2DPNE response tensors. The cross (\xmark) and the tick (\cmark) mark represents the corresponding response tensor is symmetry forbidden and allowed, respectively.}
\label{2dpne_table}
\begin{tabular}{c c c c c c c c c c c c c c c c}
\hline \hline 
\noalign{\vskip 2pt}
\rm{Longitudinal } & \rm{Transverse } & ${\mathcal P}$ & ${\mathcal T}$ & $\mathcal{P}\mathcal{T}$  &  ${\cal M}_x$ & ${\cal M}_y$ & ${\cal M}_z$ & ${\cal C}_{2x}$ & ${\cal C}_{2y}$ & ${\cal C}_{2z}$ & ${\cal C}_{3z}$ & ${\cal C}_{4z}$ & ${\cal C}_{6z}$ & ${\cal S}_{4z}$  & ${\cal S}_{6z}$ \\
\noalign{\vskip 2pt}
\hline \hline 

$\chi_{xx;x}$ & $\chi_{yx;y}$ & \cmark  & \xmark  & \xmark & \cmark &  \xmark & \xmark & \cmark & \xmark & \xmark &  \cmark  & \xmark & \xmark & \xmark & \cmark \\

$\chi_{xx;y}$ & $\chi_{yx;x}$ & \cmark  & \xmark & \xmark  & \xmark &  \cmark & \xmark & \xmark & \cmark  & \xmark & \cmark  & \xmark & \xmark & \xmark & \cmark \\

\noalign{\vskip 2pt}
\hline
\noalign{\vskip 1pt}

$\chi_{xx;xx}$, $\chi_{xx;yy}$ & $\chi_{yx;xy}$ &  \cmark & \cmark & \cmark & \cmark & \cmark & \cmark &  \cmark &  \cmark &  \cmark &  \cmark &  \cmark &  \cmark &  \cmark &  \cmark \\ 

$\chi_{xx;xy}$ & $\chi_{yx;xx}$, $\chi_{yx;yy}$ &  \cmark & \cmark & \cmark & \xmark & \xmark & \cmark & \xmark & \xmark & \cmark & \cmark & \cmark & \cmark & \cmark & \cmark \\

\noalign{\vskip 2pt}
\hline \hline
\end{tabular}
\label{table_1}
\end{table*}
%%%%%%%%%%%%%%%%%%%%%%%%%
%%%%%%%%%%%

To examine constraints from ${\cal T}$ symmetry, we use the following transformation of physical quantities under time reversal: 
${\bm j}' = -{\bm j}$, $\tau' = -\tau$, ${\bm B}' = -{\bm B}$, and ${\bm \nabla}' T = {\bm \nabla} T$. Here, the prime ($'$) denotes the time-reversed quantities. Applying these to the response expression yields, 
\[
j'_a = \frac{\tau'}{eT} \left( \chi'_{ab;c} + \chi'_{ab;cd} B'_d \right) B'_c \nabla'_b T  = -j_a~.
\]
Comparing both sides gives $\chi'_{ab;c} = -\chi_{ab;c}$ and $\chi'_{ab;cd} = \chi_{ab;cd}$. Therefore, $\chi_{ab;c}$ is a ${\cal T}$-odd third-rank axial tensor, while $\chi_{ab;cd}$ is a ${\cal T}$-even fourth-rank polar tensor. As a result, the linear-$B$ response $\chi_{ab;c}$ vanishes in nonmagnetic systems, making $\chi_{ab;cd}$ the dominant contribution in systems preserving ${\cal T}$-symmetry. 

To determine the restrictions from crystalline symmetries, we apply Neumann’s principle~\cite{newnham2005properties}. Under a crystalline symmetry operation ${\cal O}$, the response tensors transform as, 
\begin{align}
\chi_{a'b';c'} &= \eta_T \det({\cal O}) {\cal O}_{a'a} {\cal O}_{b'b} {\cal O}_{c'c} \chi_{ab;c}~, \\
\chi_{a'b';c'd'} &= {\cal O}_{a'a} {\cal O}_{b'b} {\cal O}_{c'c} {\cal O}_{d'd} \chi_{ab;cd}~.
\end{align}
Here, $\eta_T = +1$ for magnetic point group operations (${\cal O} = {\cal RT}$) and $\eta_T = -1$ for nonmagnetic ones (${\cal O} = {\cal R}$), with ${\cal R}$ denoting crystalline symmetry operation. 

To enumerate symmetry-allowed components, we assume a temperature gradient ${\bm \nabla} T$ along $\hat{\bm x}$ and an in-plane magnetic field at an angle $\phi$ from the temperature gradient direction, given by ${\bm B} = B(\cos \phi, \sin \phi)$. 
For this thermal and magnetic field configuration, the third-rank 2DPNE response tensor $\chi_{ab;c}$ has $\chi_{xx;x}$ and $\chi_{xx;y}$ as longitudinal elements , while $\chi_{yx;x}$ and $\chi_{yx;y}$ are the transverse elements. However, the fourth-rank 2DPNE response tensor $\chi_{ab;cd}$, which is symmetric in the magnetic field indices (last two indices of the response tensor), has $\chi_{xx;xx}, ~\chi_{xx;yy},$ and $ \chi_{xx;xy}$ as longitudinal elements, while $\chi_{yx;xx},~ \chi_{yx;yy},$ and $\chi_{yx;xy}$ denote the transverse elements. 
Table~\ref{2dpne_table} summarizes the symmetry-allowed and forbidden components of $\chi_{ab;c}$ and $\chi_{ab;cd}$ for a range of point group operations. A particularly notable result is that ${\cal C}_{3z}$ symmetry does not eliminate any of the allowed 2DPNE tensor components. This makes hexagonal materials such as multilayer graphene, transition metal dichalcogenides, and their twisted moire heterostructures excellent candidates for realizing 2DPNE. In the next section, we explore the angular dependence of the 2DPNE current in more detail. 

\section{Angular variation of the 2DPNE currents}

The angular dependence of the 2DPNE current with respect to the in-plane angle between ${\bm \nabla} T$ and $\bm B$ offers crucial insight into the microscopic origin and symmetry constraints of the response. We consider the coplanar configuration shown in Fig.~\ref{Fig1}, where the temperature gradient is applied along the $\hat{\bm x}$ direction and the magnetic field lies in-plane, making an angle $\phi$ with ${\bm \nabla} T$. We define the longitudinal and transverse 2DPNE currents as $j_{\parallel(\perp)}^{\rm 2DPNE} = \alpha_{\parallel(\perp)} \nabla T$, where $\alpha_{\parallel(\perp)}$ denote the 2DPNE conductivities along (perpendicular to) the temperature gradient. Using Eq.~\eqref{eq:j_cur}, we obtain the following angular dependencies of these 2DPNE responses, 
\bea \label{sigma_long}
 \alpha_{\parallel} &=& \frac{\tau B}{eT} (\chi_{xx;x} \cos\phi + \chi_{xx;y}\sin\phi) + \frac{\tau B^2}{eT} (\chi_{xx;xx} \cos^2{\phi} \nn \\
&&~+~\chi_{xx;yy} \sin^2{\phi} + \chi_{xx;xy}\sin{\phi}\cos{\phi})~, 
\eea
\bea
\alpha_{\perp} &=& \frac{\tau B}{eT} (\chi_{yx;x} \cos{\phi}+\chi_{yx;y} \sin{\phi}) + \frac{\tau B^2}{eT} (\chi_{yx;xx} \cos^2{\phi} \nn \\
&& +~\chi_{yx;yy} \sin^2{\phi} + \chi_{yx;xy}\sin{\phi}\cos{\phi})~. \label{sigma_hall}
\eea
These expressions, together with the explicit form of the 2DPNE response tensors in Table~\ref{tab1} and the symmetry constraints summarized in Table~\ref{table_1}, are the main theoretical results of this work. 

For nonmagnetic systems possessing in-plane mirror symmetry (${\cal M}_x$, ${\cal M}_y$) or two-fold rotation symmetry (${\cal C}_{2x}$, ${\cal C}_{2y}$), the leading nonzero components reduce to $\chi_{xx;xx}$ and $\chi_{xx;yy}$ for the longitudinal response and $\chi_{yx;xy}$ for the transverse one. This results in the expected angular dependencies, $\cos^2\phi$ for the longitudinal and $\sin 2\phi$ for the transverse 2DPNE conductivity. 

%%%%%%%%%%%%%%%%
\section{2DPNE responses in strained bilayer graphene}\label{sec:strBLG}
To demonstrate the emergence of 2DPNE in a realistic system, we focus on Bernal-stacked bilayer graphene (BLG), a good platform due to its experimental accessibility, gate-tunable band structure, and high sample quality. The low-energy tight-binding Hamiltonian of pristine BLG is given by~\cite{McCann_2013},
\begin{equation}
\label{hamiltonian}
    \mathcal{H}=
    \begin{pmatrix}
        \Delta & -\gamma_{0}f(k) & \gamma_{4}f(k) & -\gamma_{3}f^{*}(k)\\[6pt]
        -\gamma_{0}f^{*}(k) & \Delta & \gamma_1 & \gamma_{4}f(k)\\[6pt] \gamma_{4}f^{*}(k) & \gamma_1 & -\Delta & -\gamma_{0}f(k)\\[6pt]
        -\gamma_{3}f(k) & \gamma_{4}f^{*}(k) & -\gamma_{0}f^{*}(k) & -\Delta
    \end{pmatrix}~.
\end{equation}
Here, $\Delta$ represents the layer potential asymmetry (introduced via a vertical displacement field), and the hopping parameters are $\{\gamma_0, \gamma_1, \gamma_3, \gamma_4\} = \{3.16, 0.381, 0.385, 0.14\}~\mathrm{eV}$ (see Fig.~\ref{fig:lattice}). Among these, $\gamma_4$ denotes interlayer hopping. The structure factor is defined as $f(k)=\sum_{l=1}^{3} e^{i\bm k \cdot \bm \delta_{l}}=e^{ik_{y}a/\sqrt{3}}+2e^{-ik_{y}a/2\sqrt{3}}\cos{(k_{x}a/2)}$, where $\bm{\delta}_l$ are the nearest-neighbor vectors from A to B atoms within a layer. These are given by $\bm{\delta}_1 = (0, a/\sqrt{3})$, $\bm{\delta}_2 = (a/2, -a/(2\sqrt{3}))$, and $\bm{\delta}_3 = (-a/2, -a/(2\sqrt{3}))$, with $a = 2.46~{\rm \AA}$ being the lattice constant. 

In the presence of a perpendicular displacement field, BLG preserves time-reversal ($\mathcal{T}$) and threefold rotational ($\mathcal{C}_{3z}$) symmetries but breaks space-inversion ($\mathcal{P}$) symmetry. As discussed earlier, $\mathcal{T}$ symmetry forbids the linear-in-$B$ 2DPNE response tensor $\chi_{ab;c}$, while the breaking of $\mathcal{P}$ is essential for generating in-plane Berry curvature and orbital magnetic moment components. Thus, the leading 2DPNE contributions in pristine BLG arise from the $B^2$-dependent response tensor $\chi_{ab;cd}$. Additional symmetry constraints arise from the in-plane mirror symmetry $\mathcal{M}_x$ (armchair direction). This allows only a subset of response tensor components to be nonzero, $\chi_{xx;xx}$, $\chi_{xx;yy}$, and $\chi_{yx;xy}$ (see Table~\ref{2dpne_table}). These result in the characteristic $\alpha_{\parallel} \propto \cos^2\phi$ angular dependence in the longitudinal response and $\alpha_\perp \propto \sin2\phi$ dependence in the transverse response, as expressed in Eqs.~(\ref{sigma_long}) and~(\ref{sigma_hall}). To access additional $B^2$-dependent components, such as $\chi_{yx;xx}$, $\chi_{xx;xy}$ and $\chi_{yx;yy}$, we introduce a uniaxial strain of 1\% applied at $30^\circ$ with respect to the zigzag direction. This strain breaks the $\mathcal{M}_x$ symmetry and lifts the restrictions on the 2DPNE response tensors. We present the details of how strain is incorporated into the tight-binding model in Appendix~\ref{strain_appendics}.

%%%%%%%%%%%%%%%% fig4 %%%%%%%%%%%%%%%
\begin{figure}[t!]
    \centering
    \includegraphics[width=\linewidth]{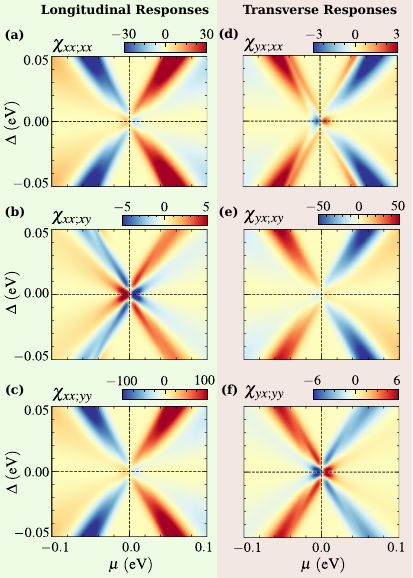}
    \caption{\textbf{Longitudinal (Seebeck) and transverse (Nernst) components of the 2DPNE responses.} 
    The non-vanishing independent components of the 2DPNE response tensor $\chi_{ab;cd}$ are plotted in the $\mu-\Delta$ space in units of nA$^2$/Tesla$^2$. The left (a-c) and right (d-f) show the longitudinal and transverse responses, respectively. We have used the same model parameters as in  Fig.~\ref{fig3} and set the system temperature to $35$~K.
    \label{fig4}}
\end{figure}

%%%%%%%%%%%%%%%%%%%%%%%%%%%%

To examine the emergence of 2DPNE in strained BLG, we first analyze its electronic structure. Figures~\ref{fig3}\tc{blue}{(a-b)} present the band dispersion and density of states (DOS). In Fig.~\ref{fig3}\tc{blue}{(a)}, the low-energy bands of strained BLG are shown as solid black lines, overlaid on the background DOS plotted as a function of energy $\varepsilon$. The band structure features saddle points near each Dirac point, giving rise to sharp peaks in the DOS or van Hove singularities (VHS). Figure~\ref{fig3}\tc{blue}{(b)} shows the evolution of these VHS across the $\mu$–$\Delta$ plane, with white dashed lines tracing the VHS. Figures~\ref{fig3}\tc{blue}{(c-d)} display the momentum-space distribution of the band dispersion for the first conduction ($c_1$) and valence ($v_1$) bands near the Dirac point. These plots show how the Fermi surface topology evolves with doping. At low chemical potential, the Fermi surface encloses the Dirac point with zero winding number. As $\mu$ increases and crosses the VHS (located near $\pm 50~\mathrm{meV}$ in our model), the winding number of the Fermi surface around the Dirac point abruptly jumps to $\pm 1$, indicating a Lifshitz transition~\cite{varlet2014anomalous,datta2024nonlinear,ahmed2025detecting}. 

To calculate the planar BC and OMM in strained BLG, we define the out-of-plane position operator $\hat{\mathcal{Z}}$. As shown in Fig.~\ref{fig:lattice}, the two graphene layers are separated by an interlayer distance $c$, and the BLG unit cell contains four carbon atoms, two per layer, each contributing one $p_z$ orbital. In this four atom basis (or layer basis) of BLG, labeled as $\{\ket{\psi^1_t}, \ket{\psi^2_t}, \ket{\psi^1_b}, \ket{\psi^2_b}\}$ for the top and bottom layers respectively, we place the origin at the mid point $z = 0$, such that the top and bottom layers lie at $z = \pm c/2$. The matrix representation of $\hat{\mathcal{Z}}$ in this orbital basis is~\cite{hara2020current,xiong2024anti,ghorai2025planar}, 
\be
\hat{\cal Z} = \frac{c}{2}{\rm diag}(1,1,-1,-1)~.
\ee 
Here, ${\rm diag}(\cdots)$ denotes a diagonal matrix with the specified elements.
Notably, the matrix representation of $\hat{\mathcal Z}$ is dependent on the choice of the basis set used to represent it. The Hamiltonian of the layered materials involves the interlayer coupling term. Therefore, we can express $\hat{\mathcal Z}$ in the eigenbasis of the Hamiltonian using the transformation: $\hat{\mathcal Z} \to \hat{U}^{\dagger} \hat{\mathcal Z} \hat{U}$. Here, $\hat{U}$ is an unitary matrix whose elements satisfies $U_{mn} = \bra{\psi_m}\psi'_{n}\rangle$ with $\ket{\psi_m}$ being the eigenstate in the layer basis, while $\ket{\psi'_n}$ denote the eigenstate of the Hamiltonian. In the eigenbasis of the Hamiltonian, $\hat{\mathcal Z}$ transforms to the off-diagonal matrix, which is crucial for computing planar BC and OMM.
Using this form of $\hat{\cal Z}$ into Eq.~\ref{eq:planar_BC_OMM}, we calculate the momentum-space distributions of the planar BC and OMM for the first conduction band. As shown in Fig.~\ref{fig3}(e), these planar band geometric quantities exhibit pronounced hotspots near the Dirac points ($K$ and $K'$). 
%highlighting their significant role in 2DPNE.

%%%%%%% figure 5%%%%%%%%%%%%%%%
\begin{figure}[t!]
    \centering
    \includegraphics[width=\linewidth]{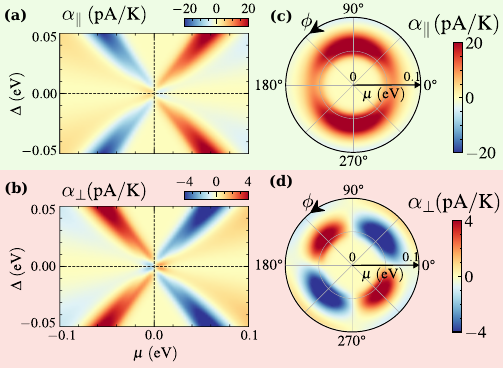}
    \caption{\textbf{Longitudinal and Transverse components of the 2DPNE conductivity for strained BLG.} (a-b) presents the variation of the longitudinal and transverse 2DPNE conductivities in the $\mu-\Delta$ space for a fixed angle between the magnetic and thermal field of $\phi = 60^\circ$. The polar plots in (c-d) display the angular variation of the 2DPNE conductivities $\alpha_{||}$ and $\alpha_{\perp}$ for a fixed interlayer potential $\Delta = 0.05$ eV with varying chemical potential. The small deviation of the responses from the conventional dependence $\alpha_{||} \propto \cos^2 \phi$ and $\alpha_{\perp} \propto \sin 2\phi$ occurs due to strain-induced mirror symmetry breaking in BLG. In these numerical computation, we considered $B = 1~$Tesla, $\tau = 1~$ps, $T = 35~$K, and the other model parameters are same as in Fig.~\ref{fig3}. 
    %\textcolor{red}{Strain is applied along txx direction.} 
    \label{fig5}}
\end{figure}
%%%%%%%%%%%%%%%%%%%

With the band structure and planar band geometric quantities established, we now examine the variation of the 2DPNE response tensor components. Figure~\ref{fig4} shows the longitudinal and transverse $\chi_{ab;cd}$ components as functions of chemical potential $\mu$ and interlayer potential $\Delta$. All components exhibit pronounced peaks near the van Hove singularities (VHS), where the density of states has a peak, and the planar BC and OMM are also prominent. These responses change sign as $\mu$ crosses the VHS and remain symmetric with respect to $\Delta$. Due to the approximate mirror symmetry $\mathcal{M}_x$ retained even with 1\% strain, components such as $\chi_{xx;xy}$, $\chi_{yx;xx}$, and $\chi_{yx;yy}$ remain significantly smaller than the dominant terms $\chi_{xx;xx}$, $\chi_{xx;yy}$, and $\chi_{yx;xy}$.

We calculate the longitudinal and transverse 2DPNE conductivities $\alpha_\parallel$ and $\alpha_\perp$ using Eqs.~(\ref{sigma_long}) and (\ref{sigma_hall}) for $\phi = 60^\circ$ and plot their dependence on $\mu$ and $\Delta$ in Figs.~\ref{fig5}(a–b). Both conductivities are sharply enhanced near the VHS and exhibit even symmetry in $\Delta$ and odd symmetry in $\mu$, consistent with the behavior of the response tensors. To highlight their angular dependence, we present polar plots of $\alpha_\parallel$ and $\alpha_\perp$ in Figs.~\ref{fig5}(c–d) for fixed $\Delta = 0.05~\text{eV}$ and varying $\mu$. These plots show clear deviations from the conventional $\cos^2\phi$ and $\sin2\phi$ forms expected in mirror-symmetric systems, reflecting the impact of strain-induced mirror symmetry breaking.

%%%%%%%%%%%

To connect our theoretical results with experiments, we estimate the induced voltage due to 2DPNE. For this, we use the following relation, 
\bea 
V_{\perp} \approx \biggl\vert \frac{j^{\rm 2DPNE}_{\perp} W}{\sigma^{D}_{||}} \biggl\vert \equiv \biggl\vert \frac{\alpha_{\perp} \nabla T_{||} W}{\sigma^{D}_{||}} \biggl\vert~.  
\eea 
Here, $W$ denotes the sample width, and $\sigma^{D}_{||}$ represents the linear Drude electrical conductivity along the direction of the applied temperature gradient. It is given by~\cite{ashcroft2022solid} $\sigma^{\rm D}_{ab} = -e^2 \tau \sum_{n, \kb} v^a_n v^b_n f'_{0,n}$. To estimate the 2DPNE voltage in strained BLG, we consider a chemical potential of $\mu = 0.05~$eV near the VHS point, a magnetic field of strength $B = 1~$Tesla, a scattering time of $\tau = 1~$ps, a relative angle between the magnetic and thermal fields of $\phi = 60^\circ$, and a system temperature of $T = 35~$K. For these parameters, from Fig.~\ref{fig5}, we obtain $\alpha_{\perp} = -4.126$ pA/K. 
We calculate the linear Drude electrical conductivity to be $\sigma^{\rm D}_{||} = -2.009~$mA/V. Assuming a sample width of $W = 10~\mu$m and a moderate temperature gradient of $\nabla T \approx 1~{\rm K}/\mu$m consistent with the experimental conditions~\cite{xu2019large}, we estimate the planar Nernst voltage as $V_{\perp} \approx 21~$nV. This lies well within the sensitivity limits of current experimental techniques~\cite{arri1978micro,aviles2018dc,avery2012observation,wu2020ac}. 
%%%%%%%%%%%%
\section{Discussion}

In this work, we focused on strained bilayer graphene (BLG), a ${\cal T}$-symmetric material where the anomalous Nernst current is strictly forbidden. However, in magnetic materials with broken time-reversal symmetry, the 2DPNE current can also be accompanied by an anomalous Nernst current (see Eq.~\eqref{eq:app_j_cur}). To isolate the 2DPNE response in experiments, we should first measure the Nernst voltage in the absence of an external magnetic field. This will quantify the baseline ANE signal (which is zero in ${\cal T}$-symmetric materials). Then, apply an in-plane magnetic field and repeat the measurement. The difference between the two provides a direct estimate of the 2DPNE signal. 

Another way to distinguish between them is to use symmetrization of the responses. Note that upon interchanging the current and temperature gradient directions (first two indices of the response tensors), the ANE response changes sign,  while the 2DPNE responses do not change sign. Thus, symmetrization of the measured response across these configurations (interchange of these indices), will only retain the 2DPNE responses, and eliminate the ANE currents. This offers another way to isolate the 2DPNE responses in experiments. 

While our analysis focuses on thermoelectric transport due to fermionic quasiparticles (electrons), thermal gradients can also drive charge-neutral bosonic excitations, such as magnons in magnetic materials and phonons in general solids. Extending the concept of planar Nernst effects to these bosonic systems represents an intriguing direction for future research.

%%%%%%%%%%
\section{Conclusion}

In this work, we introduced and systematically analyzed the two-dimensional planar Nernst effect in layered 2D materials using the  semiclassical Boltzmann transport framework. Unlike conventional PNE, which is forbidden in ideal 2D systems, we showed that interlayer tunneling in layered materials generates hidden in-plane components of Berry curvature and orbital magnetic moment which enable a robust 2DPNE response manifesting as planar Seebeck and Nernst currents. Our analysis provides a comprehensive theoretical framework and establishes the fundamental symmetry requirements for realizing this effect.

We illustrate our findings using Bernal-stacked bilayer graphene (BLG), demonstrating that 2DPNE in this system is sizable and gate-tunable. Our numerical estimates predict a thermoelectric voltage of approximately 20 nV, well within the sensitivity of current experimental techniques. We showed that the 2DPNE current exhibits strong angular dependence on the relative orientation between the applied temperature gradient and magnetic field. These findings broaden our understanding of planar thermoelectric transport in layered 2D materials and establish a clear path toward experimental realization of 2DPNE using existing measurement platforms. 

Beyond BLG, our study highlights a wide class of layered 2D systems, such as transition metal dichalcogenides and van der Waals heterostructures, as promising candidates for realizing 2DPNE. Its tunability via gate voltage, magnetic field orientation, and strain opens up exciting opportunities for thermoelectric, magnetic sensing, and energy harvesting technologies. Our work lays the foundation for further exploration of previously ignored band geometric effects in layered 2D materials and their potential applications.

%%%%%%%%%%%%
\section{Acknowledgements}

R. B. and H. V. thank Koushik Ghorai (IIT Kanpur, India) for fruitful discussions. H. V. acknowledges the Ministry of Education (MoE), government of India, for funding support through the Prime Minister's Research Fellowship. R. B. acknowledges the Indian Institute of Technology Kanpur for the Ph.D. fellowship. A. A. acknowledges funding support from the Anusandhan National Research Foundation for the Core Research Grant via Project No. xx, and the Department of Science and Technology for Project No. DST/NM/TUE/QM-6/2019 (G)-IIT Kanpur of the Government of India for funding. 
%%%%%%%%%%% appendix part
\appendix

% \onecolumngrid
%%%%%%%%%%%%%%%%%%%

\section{Planar Berry Curvature and Orbital Magnetic Moment}
\label{planar_BC}

The Berry curvature for the $n$-th Bloch band of a crystalline system is defined as\cite{xiao2010berry}~,
\begin{equation}
    \bm \Omega_{n \bm k} = \gradient_{\bm k} \times \mathcal{A}_{n k} = i \bra{\gradient_{\bm k} u_{n k}} \times  \ket{\gradient_{\bm k} u_{n k}}~.
\end{equation}
Here, $\ket{u_{nk}}$ is the periodic part of the Bloch wave function of the $n$-th band and $\mathcal{A}_{n \bm k} = \bra{u_{n\kb}}i\gradient_{\kb}\ket{u_{n\kb}}$ is the Berry-connection. We can express $\Omega_{n \bm k}$ can in terms of band velocity elements $ v_{nn'} = (1/ \hbar)\bra{u_{n \bm k}} \gradient_{\bm k} \mathcal{H}_{\bm k} \ket{u_{n' \bm k}}  $ as 
% with the system's Hamiltonian $\mathcal{H}_{k}$ as,
%
\begin{equation}
\label{Omega_3d}
    \bm{\Omega}_{n\bm k} = i\hbar^2 \sum_{n'\neq n}\frac{\bm{v}_{nn'}\times \bm{v}_{n'n}}{(\varepsilon_{n\bm k}-\varepsilon_{n'\bm k})^2}~,
\end{equation}
where $\mathcal{H}_{k}$ is the system's unperturbed Hamiltonian. In obtaining this equation, we have used the completeness relation of the Bloch wavefunction, $\sum_{n} \ket{u_{n\kb}}\bra{u_{n\kb}} = 1$, along with the interband Berry connection expression ${\mathcal A}_{nn'} = -i\hbar {\bm v}_{nn'}/(\e_{n\kb} - \e_{n'\kb})$. Here, $\e_{n\kb}$ is the energy eigenvalue of ${\mathcal H}_{\kb}$ corresponding to the Bloch state $\ket{u_{n\kb}}$. 
%Being a quantum mechanical particle, the electron is described through the wavepacket, which rotates and translates in the momentum space. 
In addition to the BC, electron wavepackets also possess an orbital magnetic moment (OMM) which is given by\cite{Chang_Niu_OMM}~,
\begin{equation}
\label{m_3d}
    \bm{m}_{n\bm k} = \frac{ie\hbar}{2} \sum_{n'\neq n}\frac{\bm{v}_{nn'}\times \bm{v}_{n'n}}{(\varepsilon_{n\bm k}-\varepsilon_{n'\bm k})}~.
\end{equation}
For a perfectly 2D system where the electron's motion is confined only in the plane, the vector product of velocity elements $\bm{v}_{nn'}\times \bm{v}_{n'n}$ will have only out-of-plane components. Consequently, from Eqs.~(\ref{Omega_3d}-\ref{m_3d}), we infer that the BC and OMM will have only out-of-plane components in perfectly 2D systems. 

However, quasi-2D materials with multiple atomic layers can host in-plane components of the BC and OMM due to finite interlayer tunneling. Assuming that the layers are stacked in the $\hat{z}$ direction, then the $z$-component of the velocity matrix can be rewritten as 
%%%%%
\begin{eqnarray}
    %\begin{aligned}
        v^{z}_{n'n} &= & \frac{1}{\hbar} \braket{ u_{n'\kb} |  \partial_{k_z} \hat{\mathcal H}_{\bm{k}} | u_{n \kb}}~ , \\ 
        &=& \frac{i}{\hbar} \braket{u_{n'\kb} | [ \hat{\mathcal H}_{\bm{k}},\hat{\cal Z}] | u_{n \kb}}~ = \frac{i}{\hbar}(\varepsilon_{n' \bm k}-\varepsilon_{n \bm k})\mathcal{Z}_{n'n}~. \nn
    %\end{aligned}
\end{eqnarray}
%%%%
Here, $\mathcal{Z}_{n'n}=\braket{u_{n'k}| \mathcal{Z} | u_{nk}}$ is the matrix element of the out-of-plane position operator $\mathcal{Z}$. In obtaining the above equation, we introduced the Heisenberg equation where $[\cdot,\cdot]$ denotes the commutator bracket. The vector product of $\bm{v}_{nn'}\times {\bm v}^z_{n'n}$ will always give an in-plane component. Introducing this description in Eqs.~(\ref{Omega_3d}-\ref{m_3d}), we deduce the planar (or in-plane) BC and OMM as 
\begin{equation}
\bm \Omega^{\rm Pl}_{nk} =2\hbar \sum_{n'\neq n}\frac{\text{Re}(\bm v_{nn'} \times \bm{\mathcal{ Z}}_{n'n})}{(\varepsilon_{n \bm k}-\varepsilon_{n' \bm k})}~,
\label{Omega_pl}
\end{equation}

\begin{equation}
\label{m_pl}
\bm m^{\rm Pl}_{nk} =e\sum_{n'\neq n}\text{Re}(\bm v_{nn'} \times \bm{\mathcal{Z}}_{n'n})~.
\end{equation}
%
%where superscript `Pl' represents the Planar component of that quantity.

%%%%%%%%%%
\section{Calculation of the 2DPNE current}
\label{2DPNE_current_density}

Here, we demonstrate how the planar band geometric quantities, the planar BC and OMM, give rise to 2DPNE. 
%In Eq.~\eqref{eq:j_trans}, we have discussed the semiclassical description for calculating thermoelectric current. Therefore, to demonstrate the planar Nernst effect within the semiclassical Boltzmann framework, 
The equation of motion of Bloch electrons for the $n$-th energy band in the presence of a temperature gradient $-\gradient T$ and external magnetic field ${\bm B}$ is given by, ~\cite{xiao2010berry,das2019berry} 
\bea 
\Dot{\bm{r}}_{n} &=& D_{\kb} \left(\tilde{\bm v}_{n\bm{k}}+\frac{e}{\hbar} \bm B(\tilde{\bm{v}}_{n\bm{k}}. \bm{\Omega_{n\bm{k}}})\right)~, \label{r_dot}\\
\hbar \Dot{\bm k}_{n} &=& -D_{\kb} e (\tilde{\bm{v}}_{n\bm{k}}\times \bm{B})~.\label{k_dot}
\eea
Here, $\tilde{\varepsilon}_{n\bm{k}}=\varepsilon_{n\bm{k}} + \e^{ m}_{n\kb}$ is the OMM modified band energy, where $\e^{ m}_{n\kb} = - \bm{m}_{n\bm{k}}\cdot{\bm{B}}$ is the linear order correction to band energy due to Zeeman-type coupling between the OMM $\bm{m}_{n\bm{k}}$ and the magnetic field ${\bm B}$. We have defined the OMM modified band velocity as, $\tilde{\bm{v}}_{n\bm{k}}=\frac{1}{\hbar}\gradient_{\bm{k}}\tilde{\varepsilon}_{\bm{k}}=\bm{v}_{n\bm{k}} + \bm{v}^{m}_{n\bm{k}}~,$ where $\bm{v}^{m}_{n\bm{k}}= (1/\hbar)\gradient_{\kb}\e^m_{n\kb} =  -({1}/{\hbar})\bm{\nabla}_k (\bm{m}_{n\bm{k}}\cdot{\bm{B}})$ is the OMM-induced velocity correction. In Eq.~\eqref{r_dot}, the 
 phase-space factor is given by $D_{\kb}=[1+{e}/{\hbar}(\bm{B}\cdot\bm {\Omega_{n\bm{k}}})]^{-1}$. Going forward, we will not explicitly write the band and momentum index appearing as the subscript $n \kb$ in various quantities for brevity. 

Next, we calculate the occupation function of Bloch electrons using the Boltzmann transport equation. Using the relaxation time approximation, we have, 
\begin{equation}
\frac{\partial f}{\partial t} + \Dot{\bm{r}} \cdot \gradient_{\bm r} f + \Dot{\bm{k}} \cdot \bm{\nabla_{k}}f= -\frac{f-\tilde{f}_0}{\tau}~.
\end{equation}
Here, $f =f( {\bm r},\kb)$ is the non-equilibrium distribution function, $\tau$ is the relaxation time which we consider to be a constant, and $\tilde{f}_{0}=(1+e^{\beta (\tilde{\varepsilon}_{n\bm{k}}-\mu)})^{-1}$ is the equilibrium Fermi-Dirac distribution corresponding to the OMM-modified band energy $\tilde{\bm{\varepsilon}}_k$ with chemical
potential $\mu$ and $\beta = 1/(k_B T)$. We calculate the steady state non-equilibrium distribution function, for which we have $\partial f/\partial t$ equal to zero. Substituting the expressions of $\dot{\bm r}$ and $\dot{\bm k}$ in the above equation, we obtain non-equilibrium distribution as
\bea
\label{B5}
    f &=& \tilde{f}_{0}+ D_{\kb}\tau\left(\tilde{\bm v}+\frac{e\bm{B}}{\hbar}(\tilde{\bm v}\cdot\bm\Omega) \right) \cdot \left( \frac{(\tilde{\varepsilon}-\mu)}{T}\gradient_{r}T
    \right)\frac{\partial f}{\partial \tilde{\varepsilon}} \nn \\ 
    &+& D_{\kb}\tau e (\tilde{\bm v} \times {\bm B}) \cdot \gradient_{\kb} f~. 
\eea
In obtaining this equation, we used the relation $\gradient_{\bm r} f = - ((\e - \mu)/T) \gradient_{\bm r} T \partial f/\partial {\tilde\e}$. For the 2D planar geometry, the last term of the above equation vanishes. We solve this self-consistent equation perturbatively by expanding $f$ into various powers of temperature gradient as $f = {\tilde f}_0 + f^{(1)} \gradient_{\bm r} T + f^{(2)} (\gradient_{\bm r} T)^2 \cdots$. For the linear response in the applied temperature gradient, we obtain, 
\be
\label{A4}
f = \tilde{f}_{0}+ D_{\kb}\tau\left(\tilde{\bm v}+\frac{e\bm{B}}{\hbar}(\tilde{\bm v}\cdot\bm\Omega) \right) \cdot \left( \frac{(\tilde{\varepsilon}-\mu)}{T}\gradient_{r}T
    \right)\frac{\partial {\tilde f}_0}{\partial \tilde{\varepsilon}}~. 
\ee
Furthermore, we will work in the low magnetic field regime, and we approximate the OMM-modified Fermi-Dirac distribution function in terms of the OMM-induced band energy correction $\e^m$ as 
$$\tilde{f}_0 \simeq f_0+ \e^m_\kb \frac{\partial f_0}{\partial \varepsilon_k} + (\e^m_{\kb})^{2}\frac{\partial^2 f_0}{\partial \varepsilon^{2}_k} + \cdots~.$$
Similarly, we expand the phase-space factor to obtain, 
$$D_{\kb}  = ( 1 + \Omega^B)^{-1}\simeq 1-\Omega^{B}+(\Omega^{B})^2 + \cdots~,$$
where we have defined, $\Omega^B = (e/\hbar) {\bm B}\cdot{\bm \Omega}$.

Using Eq.~\eqref{B5} in Eq.~\eqref{eq:j_trans}, we obtain the generalized thermoelectric current along the spatial direction $a$ as 
\bea\label{eq:app_j_cur}
&& j_a = - \frac{k_{B}e}{\hbar} \int[d\kb] \epsilon_{abc}\Omega^{c} \zeta \nabla_b T - e\tau \int  [d\kb] D_{\kb} \frac{(\tilde{\varepsilon}-\mu)}{T} \times~ \nn  \\ 
&&  \left(\tilde{ v}_{a}+\frac{e B_{a}}{\hbar}(\tilde{\bm v}.\bm\Omega) \right) \left(\tilde{ v}_{b}+\frac{eB_{b}}{\hbar}(\tilde{\bm v}.\bm\Omega) \right)\frac{\partial \tilde{f_0}}{\partial \tilde{\varepsilon}_k} \nabla_b T~. 
\eea
Here, we have defined $\zeta=\beta (\tilde{\varepsilon}-\mu) \tilde{f}_0+ \log(1+ e^{-\beta(\tilde{\varepsilon}-\mu)})$, and $\epsilon_{abc}$ denotes the third rank fully antisymmetric Levi-Civita tensor. The first term of the above equation is the anomalous Nernst current~\cite{xiao2006berry, das2019berry}. It generates a potential difference transverse to the applied temperature gradient in the absence of the external magnetic field. Being scattering time-independent, this current is intrinsic and arises from the Berry curvature. In contrast, the 2DPNE current arises from the interplay of the applied in-plane magnetic field and temperature gradient. The second term of Eq.~\eqref{eq:app_j_cur} contains the contribution from both the temperature gradient and the magnetic field, manifesting as a generalized planar Nernst response. In the absence of the magnetic field, this term reduces to the linear thermoelectric Drude current. This serves as a consistency check of our calculations. 

Restricting Eq.~\eqref{eq:app_j_cur} to quadratic order in $B$, by using the low magnetic field approximate expansions of $\tilde{f}_0$ and $D_{\kb}$, we obtain 
\begin{widetext}
\bea
j_a &=& \frac{e \tau}{T} \int  [d\kb] \left[1-\frac{e} {\hbar}(\bm{B} \cdot \Omega) + \frac{e^2} {\hbar^2}(\bm{B} \cdot \Omega)^2 \right] \left(\varepsilon_k-(\bm{m \cdot B}) - \mu\right)\left[(v_{a}-\frac{1}{\hbar} \partial_{a}({\bm m}\cdot {\bm B}))+ \frac{e B_{a}}{\hbar}(\bm v \cdot \bm\Omega)- \frac{e B_{a}}{\hbar^2}(\gradient_{\kb}({\bm m}\cdot {\bm B}) \cdot \bm\Omega) \right] \nn \\
&& \times \left[(v_{b}- \frac{1}{\hbar} \partial_b({\bm m}\cdot{\bm B}))+ \frac{e B_{b}}{\hbar}(\bm v \cdot \bm\Omega)- \frac{e B_{b}}{\hbar^2}(\gradient_{\kb}({\bm m}\cdot{\bm B}) \cdot \bm\Omega) \right] \left[-f^{\prime}_0+(\mathbf{m} \cdot \mathbf{B}) f^{''}_0 -\frac{1}{2}(\bm{m} \cdot \bm{B})^{2}f^{'''}_0 \right] \nabla_b T~.
\eea
\end{widetext}
Here, we have terminated the expansions to the second order in the applied magnetic field and defined $\partial_{a} \equiv \gradient_{k_a}$. In all the terms, the planar components of $\bm B$ or $\bm v$ couple to the co-planar components of $\bm \Omega$ or $\bm m$ via a scalar product. Thus, for the planar geometry in 2D systems, where the applied magnetic field is in the plane of the electron's motion, the above equation will have no contribution from the conventional out-of-plane BC and OMM. This establishes the significance of the hidden planar BC and OMM in determining the planar Nernst effect in 2D systems. 

The above equation can be rewritten as a sum of magnetic field-dependent contributions in the following form,  $j_a \approx (\cdots) + (\cdots)B + (\cdots)B^2$. Here, the first term is $B$-independent and this contribution arises either from the Drude thermoelectric current or the anomalous Nernst effect. This contribution does not depend on any planar band geometric quantities. In contrast, $B$-dependent contributions originate from the planar BC and OMM, giving rise to the planar Nernst current in the 2D system. Focusing only on the magnetic field dependent contributions, we can express Eq.~\eqref{eq:j_cur} as, $i.e,~j_{a} = ({\tau }/{eT}) \left( \chi_{ab;c} \nabla_bT B_{c} +  \chi_{ab;cd}\nabla_b {T} B_{c} B_{d} \right)$ and calculate the third and fourth rank 2DPNE response tensors $\chi_{ab;c}$ and $\chi_{ab;cd}$. We present the general expression of $\chi_{ab;c}$ and $\chi_{ab;cd}$ in Table~\ref{tab1}. 
%This completes the detailed derivation of the planar Nernst current in the 2D systems. 

%%%%%%%%%%%%%%%%%%%%%%%%%%%%%%%%%%%%%%%%%%%%%
\section{Strained bilayer graphene Hamiltonian}\label{strain_appendics}
 We have presented the tight-binding Hamiltonian of the unstrained BLG in Eg.~\eqref{hamiltonian}. In its unstrained form, the BLG respects the ${\cal M}_x$ mirror symmetry along the armchair direction as shown in Fig.~\ref{fig:lattice}(b). Due to the ${\cal M}_x$ symmetry, some components of the 2DPNE response tensor vanish, see Tab.~\ref{table_1}. To make these vanishing components finite, we break the ${\cal M}_x$ symmetry of BLG by applying a uniaxial strain. Application of a uniaxial strain modifies the bond length and hopping parameters of BLG and slightly distorts the lattice structure, disrupting the ${\cal M}_x$ symmetry. A uniaxial strain transforms the coordinates of the real and momentum space as~\cite{Naumis_2017_graphene, Nowack_bilayer_graphene},
\begin{equation}\label{r_strain}
    \tilde{\bm r} \approx (1+\tilde{\zeta})\bm r \hspace{0.5cm} \text{and} \hspace{0.5cm} \tilde{\bm k} \approx (1-\tilde{\zeta})^{T} \bm k ~,
\end{equation}
where $\tilde{\zeta}$ is the two-dimensional tensile strain tensor of strength $\zeta$. mathematically, it is defined  by~\cite{Pereira_2009_graphene, Choi_2010_graphene_control, sinha2022berry},
\begin{equation}
    \tilde{\zeta}=\zeta \begin{pmatrix}
        \cos^{2}\theta-\nu\sin^{2}\theta & (1+\nu)\sin{\theta}\cos{\theta} \\[4pt]
        (1+\nu)\sin{\theta}\cos{\theta} & \sin^{2}\theta-\nu\cos^{2}\theta
    \end{pmatrix}~.
\end{equation}
Here, $\theta$ is the angle between the principal strain axis and the zigzag direction, which we take as $\theta = 30^\circ$ for our calculations. The parameter $\nu \approx 0.165$ is the Poisson's ratio of graphene. Due to the strain-modified bond length and hopping parameters, the hopping integrals between the two carbon atoms get modified. We estimate these modified hopping integrals through $\tilde{\gamma}_{j}=\gamma_{j} \exp[-\tilde{\beta} ({\tilde{|\bm \delta|}}/{|\bm \delta|}-1)]$, where $\gamma_j~(j \in [0,1,3,4])$ is the hopping parameter of the unstrained BLG, and $\tilde\beta \approx 3.37$ is the electron Gr\"{u}neisen parameter. Using Eq.~\eqref{r_strain}, we compute the strain-modified carbon-carbon distance $\tilde{\bm \delta}$ between the nearest neighbor carbon atoms. For breaking the ${\cal M}_x$ symmetry of the BLG, we simultaneously apply a uniform uniaxial strain to both the layers of the Bernal-stacked BLG. To incorporate the effects of uniaxial strain in the unstrained BLG Hamiltonian, we replace $\gamma_{j}f(k) \to \Phi_j(\kb) = \sum_{p}\tilde{\gamma}^{p}_{j}\exp(i \bm k \cdot \tilde{\bm \delta}_p)$ in Eq.~\eqref{hamiltonian}. Here, the summing index $p$ runs over the appropriate bonds. The application of a uniaxial strain does not break the time-reversal symmetry of the BLG as $\bm \Phi^{*}(-\bm k)=\bm \Phi(\bm k)$. Thus, even in strained BLG, the third-rank 2DPNE response tensors $\chi_{ab;c}$ vanish.   

%%%%%%%%%%%%%%%%%% references 
\bibliography{Ref_new}
\end{document}